\documentclass[12pt]{article}

\def\be{\begin{equation}}
\def\ee{\end{equation}}
\def\bea{\begin{eqnarray}}
\def\eea{\end{eqnarray}}
\def\ba{\begin{array}}
\def\ea{\end{array}}
\def\nn{\nonumber}

\evensidemargin \oddsidemargin

\begin{document}

\newcommand{\story}{\vspace{5mm} \noindent $\spadesuit$ }

\begin{titlepage}
\vspace{15mm}
\baselineskip 9mm
\begin{center}
  {\Large \bf Hawking radiation from black holes in de Sitter spaces}
\end{center}

\baselineskip 6mm
\vspace{5mm}
\begin{center}
  Qing-Quan Jiang\footnote{E-mail address: jiangqq@iopp.ccnu.edu.cn}
 \\[4mm]
 {\it Institute of Particle Physics, Central China Normal University,
Wuhan, Hubei 430079, People's Republic of China}
\end{center}

\thispagestyle{empty}

\begin{center}
{\bf Abstract}
\end{center}
\noindent
Recently, Hawking radiation has been treated, by Robinson and Wilczek, as a compensating flux of the energy
momentum tensor required to cancel gravitational anomaly at the event horizon(EH) of a Schwarzschild-type
black hole. In this paper, motivated by their work, Hawking radiation from the event horizon(EH) and the de
Sitter cosmological horizon(CH) of black holes in de Sitter spaces, specifically including the
purely de Sitter black hole, and the static, spherically symmetric Schwarzschild-de Sitter black hole as
well as the rotating Kerr-de Sitter black hole, has been studied by anomalies. The result shows
that the gauge current and energy momentum tensor fluxes, required to restore gauge invariance and general
coordinate covariance at the EH and the CH, are precisely equal to those of Hawking radiation from the EH
and the CH, respectively. It should be noted that, gauge and gravitational anomalies taken place at the cosmological
horizon(CH) arise from the fact that the effective field theory is formulated inside the CH
to integrate out the classically irrelevant outgoing modes at the CH, which is different from those
at the black hole horizon(EH).
\\ [5mm]
\\[1mm]
\vspace{5mm}
\end{titlepage}

\baselineskip 7mm

\section{Introduction}\label{intro}

Hawking radiation from black holes, since Stephon Hawking first proved it in 1974\cite{SWH}, has been attracted a lot
of attentions by theoretical physicists. Till now, we can derive Hawking radiation from black holes in many ways such
as Euclidean quantum gravity\cite{GH}, string theory \cite{SC} and a tunnelling picture \cite{PW1, PW2, MJ} and so on.
But each method exhibits a strength and weaknesses (specifically described in Ref.\cite{RW}). Recently, Robinson and
Wilczek have proposed another method to derive Hawking radiation via the cancellation of gravitational anomaly at the
event horizon(EH) of a Schwarzschild-type black hole\cite{RW}. In the literature, the authors attempt to formulate an
effective field theory that only describes an observable physics. In this case, the global Killing vector for a static
Schwarzschild-type black hole that appears locally like a time translation, but is only time-like outside the event horizon
of the black hole could be a perfectly reasonable choice with which to define the energy of quantum states. However,
this definition would give a divergent energy at the event horizon(EH) due to the fact that the classically irrelevant
ingoing modes pile up here. To properly describe an observable physics in the effective field theory, the classically
irrelevant ingoing modes must be integrated out at the event horizon(EH) to remove the divergent energy. The effective
field theory thus formed no longer has observable divergences, but it now suffers from the destruction of general coordinate
symmetry since the number of the ingoing modes and the outgoing modes are no longer identical. Thus the effective chiral theory
contains an anomaly with respect to general coordinate symmetry, which is named as gravitational anomaly and often takes
the form of nonconservation of the energy momentum tensor. To restore general coordinate covariance at the quantum level,
an energy momentum tensor flux must be introduced to cancel gravitational anomaly at the event horizon(EH). The result shows
that the compensating energy momentum tensor flux is exactly equal to that of $(1+1)$-dimensional blackbody radiation at the
Hawking temperature. In \cite{RW}, a key technique, that each partial wave of quantum fields in an arbitrarily dimensional
black hole behaves, by a dimensional reduction, like an independent two dimensional scalar field in the near-horizon limit,
has been introduced. Subsequently, Iso etc. have extended the method to the case of a charged black hole\cite{IUW1} and a
rotating black hole\cite{IUW2} via gauge and gravitational anomalies at the black hole horizon(EH). Here, the gauge charge
of the effective two-dimensional fields is the electric charge $e$ of the radiated particles for a charged black hole, but
that for a rotating black hole is the radiated particles' azimuthal quantum number $m$.

Till now, this method has been successfully applied in many different cases\cite{ALL1, ALL2, ALL3, ALL4, ALL5, ALL6, ALL7, HJ}, but gauge
and gravitational anomalies in these observations share in common to take place at the black hole horizon(EH). In other words,
present work via anomalous point of view to derive Hawking radiation is limited to study that radiated from the black hole
horizon(EH). In fact, in de Sitter spaces, future infinity is spacelike, which means the universe with a repulsive $\Lambda$
term will expand so rapidly that for each observer there are regions from which light can never reach him. The de Sitter
cosmological horizon(CH) is the boundary of this region. Particle can be also created at such a cosmological horizon with
a thermal spectrum\cite{GH}. Although it is not of much practical significance since the temperature carried by the thermal
radiation is very small, researches on black holes in de Sitter spaces become important due to the following reasons: $(1)$
The recent observed accelerating expansion of our universe indicates the cosmological constant might be a positive one\cite{ASJ};
$(2)$Conjecture about de Sitter/conformal field theory(CFT) correspondence\cite{AD}.

Although the black hole horizon(EH) and the cosmological horizon(CH) share many similar properties, at which the radiation behavior behaves
completely different. At the black hole horizon(EH), from which an observer who lives outside the horizon would find an outgoing flux
with a thermal spectrum radiated. However, an ingoing thermal flux from the CH would be detected by an observer who lives inside the
de Sitter cosmological horizon(CH). Since the cosmological horizon(CH) is a null hypersurface, the outgoing modes that fall out of the de
Sitter cosmological horizon would never fall classically back, but Quantum mechanically its effect on the physics inside the cosmological
horizon should be taken into account. If the effective field theory is formulated to integrate out the classically irrelevant outgoing
modes at the CH, it is chiral here, but contains gauge and gravitational anomalies. To restore gauge invariance and general coordinate
covariance at the quantum level, we must introduce the gauge current and energy momentum tensor fluxes to cancel these anomalies
at the cosmological horizon. The result shows that these compensating fluxes are exactly equal to those of Hawking radiation from the de
Sitter cosmological horizon(CH).

In this paper, taking the purely de Sitter black hole, the static, spherically symmetric Schwarzschild-de Sitter black hole and
the rotating Kerr-de Sitter black hole as an example, we study Hawking radiation from the black hole horizon(EH) and the de Sitter
cosmological horizon(CH) via anomalies. For the Schwarzschild-de Sitter black hole and the Kerr-de Sitter black hole, there are a black
hole horizon(EH) and a de Sitter cosmological horizon(CH) for an observer moving on the world line of constant $r$ between the EH and
the CH. Thus, the effective field theory that describes an observable physics should be formulated between the EH and the CH to, respectively,
integrate out the classically irrelevant ingoing modes at the EH and the classically irrelevant outgoing modes at the CH. Now, gauge and
gravitational anomalies take place at both the EH and the CH. To simplify our discussion, when dealing with Hawking radiation from the EH,
we can regard that gauge and gravitational anomalies taken place in the effective field theory are due to integrate out the classically irrelevant
ingoing modes at the EH, and disregard the quantum contribution of the omitted outgoing modes at the CH although they should be incorporated
into the effective field theory. This means we have overlooked the effect coming from the CH when we study Hawking radiation from the EH,
and assumed that the EH and the CH behave like two independent systems. In the CH case, we can similarly treat gauge and gravitational
anomalies occurred in the effective field theory as arising from excluding the classically irrelevant outgoing modes at the CH, and overlook
the effect of the EH. This dealing has been successfully applied in Ref.\cite{MJ} to derive Hawking radiation via tunnelling from black holes in
de Sitter spaces. The result shows that although gauge and gravitational anomalies take place in different ways at the black hole
horizon(EH) and the cosmological horizon(CH), the gauge current and energy momentum tensor fluxes, required to restore gauge invariance
and general coordinate covariance at the quantum level, are always equal to those of Hawking radiation.

The organization of the paper is as follows. In Sec.\ref{sbh}, by extending the Robinson-Wilczek's work that Hawking radiation
can be determined by anomaly cancellation at the black hole horizon(EH), we study Hawking radiation from the de Sitter cosmological
horizon(CH) of the purely de Sitter black hole. Sec.\ref{rbh} and Sec.\ref{ks} are, respectively, devoted to investigating Hawking
radiation from the EH and the CH of the static, spherically symmetric Schwarzschild-de Sitter black hole and the rotating Kerr-de Sitter
black hole via gauge and gravitational anomalies. Sec.\ref{cm} endes up with conclusions and discussions.

\section{Hawking radiation from the de Sitter black hole }\label{sbh}
In de Sitter spaces, the simplest solution for the Einstein field equation with $T_{ab}=0$ is written as
\be
ds^2=f(r)dt^2-\frac{1}{f(r)}dr^2-r^2(d\theta^2+\sin^2\theta d\phi^2), \label{m}
\ee
where
\be
f(r)=1-\frac{1}{3}\Lambda r^2.
\ee
Obviously, the cosmological horizon of the solution is located at $r_c=\sqrt{3}/\sqrt{\Lambda}$, and the
surface gravity for the horizon is given by $\kappa_c=-\frac{1}{2}\partial_r f(r)|_{r=r_c}=\sqrt{\Lambda}/\sqrt{3}$.
Before deriving Hawking radiation via gravitational anomaly at the horizon, we must introduce a dimensional
reduction technique to reduce the scalar field theory on this metric to an effective two-dimensional theory. For the
metric described by Eq.(\ref{m}), the action of the scalar field reads off
\bea
S[\varphi]&=&\frac{1}{2}\int d^4 x\sqrt{-g} \varphi \nabla^2\varphi \nn\\
&=&\frac{1}{2}\int d^4x r^2\sin\theta \varphi \Big[\frac{1}{f(r)}\partial_t^2-\frac{1}{r^2}\partial_r r^2f(r)\partial_r \nn\\
&-&\frac{1}{r^2}\Big(\partial_\theta^2+\frac{1}{\sin^2\theta}\partial_\phi^2\Big)\Big]\varphi.
\eea
Near the horizon, performed the partial wave decomposition of $\varphi$ in terms of the spherical harmonics $\varphi=\sum_n \varphi_n
(t,r) Y_n(\theta,\phi)$, the action then becomes
\be
S[\varphi]=\sum_n\frac{r^2}{2}\int dtdr \varphi_n\Big(\frac{1}{f(r)}\partial_t^2-\partial_r f(r)\partial_r\Big)\varphi_n.
\ee
Now each partial wave for the four dimensional de Sitter black hole can be effectively described by an infinite two-dimensional
scalar field on the metric
\be
ds^2=f(r)dt^2-\frac{1}{f(r)}dr^2, \label{s}
\ee
and the dilaton background $\Psi=r^2$ whose contributions are often dropped. In the two-dimensional reduction, the energy momentum tensor
is conserved due to the static background, which generates a general coordinate symmetry for the two-dimensional scalar field.
When the effective field theory is formulated inside the cosmological horizon(CH) to integrate out the classically irrelevant
outgoing modes at the CH, it is chiral here since an observer who lives inside the cosmological horizon no longer detect a divergent
energy at the CH, but it now suffers an anomaly with respect to general coordinate symmetry, which gives a great constraint on the
energy momentum tensor. For the left-handed field(ingoing modes), the consistent anomaly reads\cite{LE}
\be
\nabla_\mu T_\nu^\mu= -\frac{1}{\sqrt{-g}}\partial_\mu \mathcal{N}_\nu^\mu, \label{T}
\ee
where
\be
\mathcal{N}_\nu^\mu=\frac{1}{96\pi}\epsilon^{\beta\mu}\partial_\alpha\Gamma_{\nu \beta}^\alpha, \label{N}
\ee
and the two-dimensional Levi-Civita tensor $\epsilon^{01}=1$. In the effective field theory, the total energy momentum tensor
combines contributions from two regions $T_\nu^\mu=T_{\nu(o)}^\mu\Theta_-+T_{\nu(C)}^\mu C$, where $\Theta_-=\Theta(r_c-r-\epsilon)$ and
$C=1-\Theta_-$ are, respectively, a scalar step function and scalar top hat function. Near the cosmological horizon $r_c-\epsilon\leq r \leq r_c$,
gravitational anomaly take place in the effective field theory due to integrating out the classically irrelevant outgoing modes here, the
nonconservation of the energy momentum tensor in this region satisfies $\partial _rT_{t(C)}^r=-\partial_r\mathcal{N}_t^r(r)$ (here we only
consider the $\nu=t$ component since the anomaly is purely time-like), where
\be
\mathcal{N}_t^r(r)=\frac{1}{192\pi}(f'^2+ff'') \label{N1}.
\ee
In the other region $r\leq r_c-\epsilon$, there is no anomaly, and the energy momentum tensor satisfies the conservation law as
$\partial_r T_{t(o)}^r=0$. In classical theory, the covariance of the effective action under general coordinate transformation is
expressed as $\delta_\lambda W=-\int d^d x\sqrt{-g}\lambda^\nu \nabla_\mu T_\nu^\mu=0$ where $\lambda$ is the variational parameter.
In our cases, the effective field theory is formulated inside the cosmological horizon to integrate out the classically irrelevant
outgoing modes at the cosmological horizon(CH). Under general coordinate transformation, the effective action(without incorporating the
quantum contributions of the classically irrelevant outgoing modes at the CH) changes as
\bea
-\delta_\lambda W &=& \int dtdr\sqrt{-g}\lambda^t\nabla_\mu \big(T_{t(o)}^\mu\Theta_-+T_{t(C)}^\mu C\big) \nn\\
&=& \int dtdr \lambda^t\Big[\big(T_{t(C)}^r-T_{t(o)}^r+\mathcal{N}_t^r\big)\delta(r-r_c+\epsilon) \nn\\
&-&\partial_r(C \mathcal{N}_t^r)\Big]. \label{w}
\eea
In Eq.(\ref{w}), the the second term should be cancelled by the quantum effect of the classically irrelevant outgoing modes at the CH,
whose contributions to the total energy momentum tensor are $C \mathcal{N}_t^r$. To restore general coordinate covariance at the quantum level,
the coefficient of the delta function should also vanish, which means
\be
a_o=a_c+\mathcal{N}_t^r(r_c),
\ee
where $a_o=T_{t(o)}^r$ is the energy flow observed by an observer who lives inside the cosmological horizon(CH), and
\be
a_c=T_{t(C)}^r+\int_{r_c}^r dr \partial_r  \mathcal{N}_t^r,
\ee
are the value of the energy flow at the cosmological horizon(CH). To ensure the regularity of the physics quantities
at the CH, the covariant energy momentum tensor, which is related to the consistent one by \cite{IUW1}
\be
\widetilde{T}_{t(C)}^r=T_{t(C)}^r-\frac{1}{192\pi}(ff''-2f'^2), \label{11}
\ee
should vanish here. This regular condition determines the energy momentum tensor flux at the CH should be taken the form as
\be
a_c=-\frac{\kappa_c^2}{24\pi}=-2\mathcal{N}_t^r(r_c).
\ee
Thus, to cancel gravitational anomaly at the cosmological horizon(CH), the total energy momentum tensor flux is given by
\be
a_o=-\mathcal{N}_t^r(r_c)=-\frac{\pi}{12}T_c^2, \label{ao2}
\ee
where
\be
T_c=\frac{\kappa_c}{2\pi}=\frac{\sqrt{\lambda}}{2\sqrt{3}\pi},
\ee
is the Hawking temperature at the cosmological horizon of the black hole. In (\ref{ao2}), the negative sign denotes that the effective field theory
would absorb the energy momentum tensor flux to ensure general coordinate covariance at the quantum level. In fact, this absorbing energy momentum
tensor flux is exactly equal to that of Hawking radiation from the CH. For fermions, the Hawking distribution at the de Sitter cosmological
horizon(CH) of the black hole is given by $\mathcal{N}(\omega)=-1/[\exp(\frac{\omega}{T_c})+1]$ (here the negative sign denotes blackbody radiation
is radiated into the black hole from the CH). With this distribution, the energy momentum tensor flux is given by
\be
F_c=\int_0^\infty \frac{\omega}{\pi}\mathcal{N}(\omega)d\omega=-\frac{\pi}{12}T_c^2.
\ee
Obviously, the total energy momentum tensor flux, required to cancel gravitational anomaly at the cosmological horizon(CH) and restore
general coordinate covariance at the quantum level, has an equivalent form as that of Hawking radiation from the CH.

In the following section, we will further extend the Robinson-Wilczek's method to study Hawking radiation from the event horizon(EH) and
the de Sitter cosmological horizon(CH) of the Schwarzschild-de Sitter black hole. In the two-dimensional reduction, if the effective field
theory is formulated between the EH and the CH to, respectively, exclude the classically irrelevant ingoing modes at the EH and the
classically irrelevant outgoing modes at the CH, gravitational anomaly would take place at both the EH and the CH. In our discussion, we
take the simplest case as discussed above to derive Hawking fluxes from black holes in de Sitter spaces.

\section{Hawking radiation from the Schwarzschild-de Sitter black hole}\label{rbh}

The Schwarzschild solution with a repulsive constant $\Lambda$ represents a black hole in asymptotically de Sitter space.
The metric of a four-dimensional Schwarzschild-de Sitter black hole can be written as
\be
ds^2=f(r)dt^2-\frac{1}{f(r)}dr^2-r^2(d\theta^2+\sin^2\theta d\phi^2),
\ee
where
\be
f(r)=1-\frac{2M}{r}-\frac{1}{3}\Lambda r^2 \label{f}.
\ee
If $\Lambda>0$ and $9\Lambda M^2<1$, $f(r)$ is zero at the two positive values of $r$. In which, the smaller positive one, denoted by $r_h$
can be regarded as the position of the black hole horizon, while the larger value $r_c$ represents the position of the de Sitter cosmological
horizon, and $r_-$ is the negative root of $f(r)=0$. The surface gravities on the black hole horizon(EH) and the cosmological horizon(CH)
read as
\bea
\kappa_h=\frac{1}{2}\partial_r f(r)|_{r=r_h}=\frac{\Lambda}{6r_h}(r_c-r_h)(r_h-r_-), \nn\\
\kappa_c=-\frac{1}{2}\partial_r f(r)|_{r=r_c}=\frac{\Lambda}{6r_c}(r_c-r_h)(r_c-r_-),
\eea
respectively. After a dimensional reduction technique near the EH or the CH, the scalar field theory in the original dimensions
can always be treated as an infinite collection of $(1+1)$-dimensional fields on the metric described by Eq.(\ref{s}) only replacing
$f(r)$ with that in Eq.(\ref{f}). In this section, Hawking radiation from both the EH and the CH will be derived from anomalous point of view. Now
we first study the energy momentum tensor flux and gravitational anomaly at the EH.

\subsection{Hawking radiation from the EH}

In the Schwarzschild-de Sitter spacetime, the surface $r=r_h$ and $r=r_c$ are, respectively, the black hole horizon(EH) and the de Sitter
cosmological horizon(CH) for an observer moving on the word lines of constant $r$ between the EH and the CH. In the two-dimensional reduction,
the effective field theory that only describes an observable physics should be then formulated between the event horizon(EH) and the
de Sitter cosmological horizon(CH) to, respectively, integrate out the ingoing modes at the EH and the outgoing modes at the CH.
Now, gravitational anomaly takes place at both the EH and the CH. When dealing with Hawking radiation from the EH, we assume
gravitational anomaly taken place in the effective field theory is due to exclude the classically irrelevant ingoing modes at
the EH, and disregard the quantum contribution of the omitted outgoing modes at the CH although they should be incorporated into
the effective field theory. Near the EH  $r_h\leq r \leq r_h+\epsilon$, the nonconservation of the energy momentum tensor then
satisfies the anomalous equation as
\be
\nabla_\mu T_{\nu(H)}^\mu= \frac{1}{\sqrt{-g}}\partial_\mu \mathcal{N}_\nu^\mu,
\ee
for right-handed fields(outgoing modes), where $\mathcal{N}_\nu^\mu$ has a completely same form as Eq.(\ref{N}). In the other region
$r_h+\epsilon\leq r \leq r_c$, without any anomalies, the energy momentum tensor satisfies the conservation equation as $\nabla_\mu T_
{\nu(o)}^\mu=0$. As the anomaly is purely time-like, the anomalous energy momentum tensor near the EH satisfies $\partial_r T_{t(H)}^r
=\partial_r \mathcal{N}_t^r$, where $\mathcal{N}_t^r$ is given by Eq.(\ref{N1}) only replacing $f(r)$ with that in Eq.(\ref{f}), and the
conservation energy momentum tensor in the other region satisfies $\partial_r T_{t(o)}^r=0$. Under general coordinate transformation, the
effective action changes as
\bea
-\delta_\lambda W &=& \int dtdr \sqrt{-g} \lambda^t \nabla_\mu (T_{t(o)}^\mu \Theta_++T_{t(H)}^\mu H)\nn\\
&=&  \int dtdr \lambda^t \Big[(T_{t(o)}^r-T_{t(H)}^r+\mathcal{N}_t^r)\delta(r-r_h-\epsilon) \nn\\
&+& \partial_r (H \mathcal{N}_t^r)\Big], \label{W2}
\eea
where $\Theta_+=\Theta(r-r_h-\epsilon)$, and $H=1-\Theta_+$ are, respectively, scalar step function and top hat function. In (\ref{W2}),
the classically irrelevant ingoing modes at the EH have been integrated out, so the second term is cancelled by its quantum effect,
whose contributions to the total energy momentum tensor flux is $-H \mathcal{N}_t^r$. To restore general coordinate covariance at
the quantum level, the coefficient of the delta function should be also vanished. This relates
\be
d_o=d_h-\mathcal{N}_t^r(r_h),
\ee
where $d_o=T_{t(o)}^r$ is the energy flow observed by an observer who lives between the EH and the CH, and
\be
d_h=T_{t(H)}^r- \int_{r_h}^r dr \partial_r \mathcal{N}_t^r,
\ee
is the energy flow at the EH. To determine the total energy momentum tensor flux, we impose the covariant energy momentum tensor is vanished
at the EH, which corresponds to the regularity condition of the physical quantities. Since the covariant energy momentum tensor is written
in the consistent one by\cite{IUW1}
\be
\widetilde{T}_{t(H)}^r=T_{t(H)}^r+\frac{1}{192\pi}[ff''-2(f')^2], \label{T7}
\ee
the vanishing condition determines the energy momentum tensor flux at the EH
\be
d_h=\frac{\kappa_h^2}{24\pi}=2\mathcal{N}_t^r(r_h).
\ee
Thus, the compensating energy momentum tensor flux, required to cancel gravitational anomaly and restore general coordinate covariance at
the quantum level, is given by
\be
d_o=\mathcal{N}_t^r(r_h)=\frac{\pi}{12}T_h^2, \label{do}
\ee
where $T_h=\kappa_h/(2\pi)$ is the Hawking temperature at the event horizon(EH) of the black hole. At the EH, as the
Planckian distribution of blackbody radiation moving in the positive $r$ direction at the Hawking temperature $T_h$ takes the form as
$ \mathcal{N}(\omega)=1/[\exp(\frac{\omega}{T_h})+1]$ for fermions, the energy momentum tensor flux with this distribution is
calculated as
\be
F_h=\int_0^\infty \frac{\omega}{\pi}\mathcal{N}(\omega)d\omega=\frac{\pi}{12}T_h^2. \label{Fh}
\ee
Compared the energy momentum tensor flux derived from the cancellation condition of gravitational anomaly at the EH with that from Hawking
radiation with the Planckian distribution at the Hawking temperature $T_h$ shows that Hawking radiation is capable of restoring general coordinate
covariance at the quantum level.

\subsection{Hawking radiation from the CH}

The previous subsection tells us that the compensating energy momentum tensor flux at the EH, required to cancel gravitational anomaly
and restore general coordinate covariance at the quantum level, is exactly equal to that of $(1+1)$-dimensional blackbody radiation
at the Hawking temperature. The simplest case, that gravitational anomaly in the effective field theory only occurs at the EH
and the quantum contributions of the omitted outgoing modes at the CH is out of consideration, has been adopted in our discussion.
In this subsection, we will concentrate on studying Hawking radiation from the cosmological horizon(CH) of the black hole via
anomalous point of view. Similarly, we can think that gravitational anomaly taken place in the effective field theory is
due to exclude the classically irrelevant outgoing modes at the CH, and the quantum effect of the omitted ingoing modes at the EH
is out of consideration. Thus, the energy momentum tensor in the effective field theory is contributed as $T_\nu^\mu=T_{\nu(o)}^\mu
 \Theta_-+T_{\nu(C)}^\mu C$, where $\Theta_- =\Theta(r_c-r-\epsilon)$ is a scalar step function, and $C=1-\Theta_-$ denotes a
scalar top hat function. According to Sec.\ref{sbh}, gravitational anomaly for left-handed fields(ingoing modes) gives a great constraint
on the consistent energy momentum tensor as described by Eq.(\ref{T}). Under general coordinate transformation, the variation of the
effective action is given by
\bea
-\delta W&=&\int dtdr\sqrt{-g}\lambda^t\nabla_\mu \big(T_{t(o)}^\mu\Theta_-+T_{t(C)}^\mu C\big) \nn\\
&=&  \int dtdr \lambda^t\Big[\big(T_{t(C)}^r-T_{t(o)}^r+\mathcal{N}_t^r\big)\delta(r-r_c+\epsilon) \nn\\
&-&\partial_r(C \mathcal{N}_t^r)\Big]. \label{w2}
\eea
In Eq.(\ref{w2}), the classically irrelevant outgoing modes at the CH has been integrated out. As the original underlying theory is,
of course, covariant and the last term should be cancelled by the quantum effect of the classically irrelevant outgoing modes at the CH, whose
contributions to the total energy momentum tensor are $C \mathcal{N}_t^r$. To restore diffeomorphism covariance at the quantum level,
the coefficient of the delta function should also vanish, which relates
\be
f_o=f_c+\mathcal{N}_t^r(r_c),
\ee
where $f_o=T_{t(o)}^r$ is the flux of the energy momentum tensor observed by an observer who lives between the EH and the CH, and
\be
f_c=T_{t(C)}^r+\int_{r_c}^r dr \partial_r \mathcal{N}_t^r,
\ee
is the energy flow at the CH. The covariant energy momentum tensor, which is related to the consistent one as Eq.(\ref{11}),
should be vanished to assure regularity of the physical quantities at the CH. That condition determines the flux of the energy
momentum tensor at the CH as
\be
f_c=-\frac{\kappa_c^2}{24\pi}=-2\mathcal{N}_t^r(r_c),
\ee
where $\kappa_c$ is the surface gravity at the de Sitter cosmological horizon(CH) of the black hole. Now the total flux of the energy momentum tensor,
required to cancel gravitational anomaly at the CH, is written as
\be
f_o=-\mathcal{N}_t^r(r_c)=-\frac{\pi}{12}T_c^2, \label{fo}
\ee
where $T_c=\kappa_c/(2\pi)$ is the Hawking temperature at the CH of the black hole, and the negative sign denotes the effective field theory must
absorb the energy flow to cancel gravitational anomaly at the CH. In fact, the energy momentum tensor flux of Hawking radiation at the CH of the
black hole has a same form as that derived from the vanishing condition of gravitational anomaly at the CH. For fermions, the Hawking distribution
at the CH of the black hole should be taken the form as $ \mathcal{N}(\omega)=-1/[\exp(\frac{\omega}{T_c})+1]$ (the negative sign denotes blackbody
radiation at the CH of the black hole is moving in the negative $r$ direction), where $T_c$ is the Hawking temperature at the CH. Integrating the
Hawking distribution, we obtain the flux of the energy momentum tensor
\be
F_c=\int_0^\infty \frac{\omega}{\pi}\mathcal{N}(\omega)d\omega=-\frac{\pi}{12}T_c^2.\label{Fo}
\ee
Obviously, the total energy momentum tensor flux required to cancel gravitational anomaly at the CH of the black hole is precisely equal to
that of Hawking radiation from the CH.

In a word, to restore general coordinate covariance at the quantum level, each partial wave of the scalar field at the EH and the CH must be, respectively,
in a state with the energy momentum tensor flux given by Eq.(\ref{do}) and Eq.(\ref{fo}). In addition, these energy momentum tensor fluxes
are exactly equal to those of $(1+1)$-dimensional blackbody radiation .

The following section is devoted to investigating Hawking radiation from the EH and the CH of the rotating Kerr-de Sitter black hole via gauge and
gravitational anomalies. In this background, there are also the black hole horizon(EH) and the de Sitter cosmological horizon(CH) for an observer
who lives between the EH and the CH. After a dimensional reduction technique, the effective two-dimensional theory for each partial wave
exhibits an $U(1)$ gauge symmetry, which originates from the isometry of the black hole along $\phi$ direction, and a general coordinate
symmetry. When the effective field theory is formulated between the EH and the CH to, respectively, integrate out the classically irrelevant
ingoing modes at the EH and the classically irrelevant outgoing modes at the CH, $U(1)$ gauge and gravitational anomalies take place. It is
expected that the total $U(1)$ gauge current and energy momentum tensor fluxes, which are required to cancel these anomalies, are precisely
equal to those of Hawking radiation.

\section{Hawking radiation from the Kerr-de Sitter black hole}\label{ks}

The Kerr-de Sitter black hole can be expressed in Boyer-Lindquist coordinates as\cite{GU}
\bea
ds^2&=&\frac{\Delta}{\rho^2}\Big(dt-\frac{a}{\Xi}\sin^2\theta d\phi\Big)^2-\rho^2\Big(\frac{dr^2}{\Delta}+\frac{d\theta^2}{\Delta_\theta}\Big)\nn\\
&-& \frac{\Delta_\theta \sin^2 \theta}{\rho^2}\Big(a dt-\frac{r^2+a^2}{\Xi} d\phi\Big)^2, \label{qq}
\eea
where
\bea
&\rho^2&=r^2+a^2\cos^2\theta, ~~~~\Xi=1+\frac{a^2}{l^2}, \nn\\
&\Delta& = (r^2+a^2)\Big(1-\frac{r^2}{l^2}\Big)-2mr,\nn\\
&\Delta_\theta& = 1+\frac{a^2}{l^2}\cos^2\theta, ~~~~~\frac{1}{l^2}=\frac{\Lambda}{3}.
\eea
Here $m$ and $a$ are the mass and rotational parameters, respectively. $\Lambda$ is the cosmological constant parameter. The event horizon(EH) ($r_+$)
and the de Sitter cosmological horizon(CH) ($r_c$) are given by the equation $\Delta=0$. To derive Hawking radiation from the black hole via anomalous point of view,
we must first introduce a dimensional reduction technique to reduce the higher-dimensional theory to the effective two-dimensional one. Ref.\cite{ALL7}
tells us, performing the partial wave decomposition of the scalar field in terms of the spherical harmonics as $\varphi = \sum_{l, m} \varphi_{lm}(t, r)
Y_{lm}(\theta, \phi)$ and transforming to the tortoise coordinate defined by $dr_*/dr = (r^2+a^2)/\Delta \equiv 1/f(r)$, one can easily observe, near the
EH or the CH, each partial wave of the scalar field $\varphi$ in the four-dimensional Kerr-de Sitter black hole can be effectively described by an
infinite collection complex scalar field in the background of a $(1+1)$-dimensional metric, the dilaton $\Psi$, and the $U(1)$ gauge field $\mathcal{A}_{\mu}$ as
\bea
g_{tt} &=& f(r) = \frac{\Delta}{r^2+a^2} \, , \qquad  g_{rr}= -\frac{1}{f(r)}  \, , \nn \\
\Psi &=& \frac{r^2 +a^2}{\Xi}, ~~\mathcal{A}_t = -\frac{\Xi a}{r^2 +a^2},  ~~ \mathcal{A}_r = 0 \, , \label{g}
\eea
where the gauge charge for the $U(1)$ gauge field is the azimuthal quantum number $m$. The specific dimensional reduction procedure can be referred to
Ref.\cite{ALL7}. Now the physics near the horizon of the black hole can be effectively described by the effective two-dimensional theory.

In the Kerr-de Sitter spacetime, there are two event horizons, namely the black hole horizon(EH) and the de Sitter cosmological horizon(CH), for an
observer who lives between the EH and the CH. The $U(1)$ gauge and gravitational anomalies take place in the effective field theory due to, respectively,
integrate out the classically irrelevant ingoing modes at the EH and the classically irrelevant outgoing modes at the CH. In the following subsections,
Hawking radiation from the EH and the CH will be discussed via gauge and gravitational anomalies. To simplify our discussion, when dealing with Hawking
radiation from the EH, we think gauge and gravitational anomalies only take place at the EH due to exclude the classically irrelevant ingoing modes,
and disregard the quantum contributions of the omitted outgoing modes at the CH. That is to say, we treat the EH and the CH as two independent physical
systems with their probably interactions being overlooked.

\subsection{Hawking radiation from the EH}

As before, the first thing is to study the $U(1)$ gauge current flux and the $U(1)$ gauge anomaly at the EH. When the effective field theory is
formulated to integrate out the classically irrelevant ingoing modes at the EH, but ignore the quantum contributions of the omitted outgoing modes
at the CH, the $U(1)$ gauge current becomes anomalous at the EH. Near the black hole horizon(EH) $r_+\leq r \leq r_++\epsilon$, it satisfies the
anomalous equation as $\partial_r \mathcal{J}_{(H)}^r={m^2}\partial_r \mathcal{A}_t/({4\pi})$. In the other region $r_++\epsilon \leq r \leq r_c$,
without any anomalies, the $U(1)$ gauge current obeys the conservation equation $\partial_r \mathcal{J}_{(o)}^r=0$. Under gauge transformation, the
effective action changes as
\bea
-\delta W &=& \int dt dr \lambda \nabla_\mu \Big(\mathcal{J}_{(H)}^\mu H+\mathcal{J}_{(o)}^\mu \Theta_+\Big) \nn\\
&=&\int dtdr \lambda \Big[\Big(\mathcal{J}_{(o)}^r-\mathcal{J}_{(H)}^r+\frac{m^2}{4\pi}\mathcal{A}_t\Big)\delta(r-r_+-\epsilon) \nn\\
&+& \partial_r\Big(\frac{m^2}{4\pi}\mathcal{A}_t H\Big)\Big]. \label{WJ}
\eea
Here, $\Theta_+(r)= \Theta(r-r_+-\epsilon)$ and $H(r)=1-\Theta_+(r)$ are the scalar step and top hat functions, respectively. In Eq.(\ref{WJ}),
the second term would be cancelled by the quantum effect of the classically irrelevant ingoing modes at the EH, whose contributions to the total
gauge current are $-{m^2} \mathcal{A}_t H/({4\pi})$. To demand the full quantum theory gauge invariance at the quantum level, the coefficient
of the delta functions should be nullified, which relates
\be
g_o=g_h-\frac{m^2}{4\pi}\mathcal{A}_t(r_+). \label{goh}
\ee
In (\ref{goh}), $g_o=\mathcal{J}_{(o)}^r$ is the gauge current flux observed by an observer who lives between the EH and the CH, and
\be
g_h=\mathcal{J}_{(H)}^r-\frac{m^2}{4\pi}\int_{r_+}^r dr \partial_r \mathcal{A}_t,
\ee
is the value of the gauge current at the EH. In order to fix the total $U(1)$ gauge current flux, we impose a constraint that the covariant
gauge current related to the consistent one by $\widetilde{J}_{(H)}^r= \mathcal{J}_{(H)}^r+{m^2}\mathcal{A}_t H/({4\pi})$ is vanished at the EH,
that condition reads
\be
g_o=-\frac{m^2}{2\pi}\mathcal{A}_t(r_+)=\frac{\Xi m^2 a}{2\pi(r_+^2+a^2)}. \label{go}
\ee
This is the compensating gauge current flux to cancel the $U(1)$ gauge anomaly at the EH, which in fact, corresponds to the angular
momentum flow of Hawking radiation from the EH of the black hole.

Next, we focus on studying the energy momentum tensor flux radiated from the EH. When excluding the classically irrelevant ingoing modes at the
EH, the total energy momentum tensor for the effective field theory contains two contributions, that is $T_\nu^\mu=T_{\nu(o)} ^\mu \Theta_+(r)+
T_{\nu(H)}^\mu H(r)$. Here, $T_{\nu(o)}^\mu$ satisfies the conservation law of the energy momentum tensor in the gauge field background
$\partial_r T_{t(o)}^r=g_o\partial_r \mathcal{A}_t$, while the other component obeys the anomalous equation\cite{IUW1}
\be
\partial_rT_{t(H)}^r=\mathcal{J}_{(H)}^r\partial_r\mathcal{A}_t+\mathcal{A}_t\partial_r\mathcal{J}_{(H)}^r+\partial_r \mathcal{N}_t^r,
\ee
where $\mathcal{N}_t^r$ takes the same form as Eq.(\ref{N1}) only replacing $f(r)$ with that in Eq.(\ref{g}). Under general coordinate transformations, the
effective action changes as
\bea
-\delta W&=& \int dtdr \lambda^t \Big[g_o\partial_r\mathcal{A}_t+\partial_r\Big(\frac{m^2}{4\pi}\mathcal{A}_t^2+\mathcal{N}_t^r\Big)H \nn\\
&+&\Big(T_{t(o)}^r-T_{t(H)}^r+\frac{m^2}{4\pi}\mathcal{A}_t^2+\mathcal{N}_t^r\Big)\delta(r-r_+-\epsilon)\Big].  \label{W5}
\eea
In (\ref{W5}), the first item is the classical effect of the background gauge field for constant current flow, and the second term is cancelled by the
quantum effect of the classically irrelevant ingoing modes at the EH, whose contributions to the total energy momentum tensor are $-\Big({m^2}
\mathcal{A}_t^2/({4\pi})+\mathcal{N}_t^r\Big)H$. To demand the effective action general coordinate covariance at the quantum level,
the coefficient of the delta function should also vanish at the EH, which relates
\be
k_o=k_h+\frac{m^2}{4\pi}\mathcal{A}_t^2(r_+)-\mathcal{N}_t^r(r_+),
\ee
where
\bea
k_o&=&T_{t(o)}^r-g_o\mathcal{A}_t(r), \nn\\
k_h &=&T_{t(H)}^r-\int_{r_+}^{r}dr\partial_r\Big[g_o\mathcal{A}_t+\frac{m^2}{4\pi}\mathcal{A}_t^2+\mathcal{N}_t^r\Big],
\eea
are, respectively, the energy flow observed by an observer who lives between the EH and the CH, and that at the EH. Taking the form of the
covariant energy momentum tensor as Eq.(\ref{T7}), and further imposing the vanishing condition on it, the total flux of the energy momentum
tensor is given by
\be
k_o=\frac{\Xi^2 m^2 a^2}{4\pi(r_+^2+a^2)^2}+\frac{\pi}{12}T_+^2, \label{ko}
\ee
where $T_+=\partial_r f(r)/(4\pi)|_{r=r_+}$ is the Hawking temperature at the EH of the black hole. In fact, the $U(1)$ gauge current and
energy momentum tensor fluxes, derived from the vanishing condition of the $U(1)$ gauge and gravitational anomalies at the EH, are exactly equal to
those of Hawking radiation with the Planckian distribution $\mathcal{N}_{\pm m}(\omega)=1/[\exp(\frac{\omega\pm m \mathcal{A}_t(r_+)}{T_+})+1]$. So
Hawking radiation from the EH of the black hole can be effectively determined by anomalous point of view.

Different from the radiation behavior at the EH, Hawking radiation is radiated into the black hole from the CH. In anomalous point of view,
the formulated effective theory must absorb the gauge current and energy momentum tensor fluxes to cancel gauge and gravitational anomalies at
the CH. We expect that these absorbing fluxes are precisely equal to those of Hawking radiation. Here, we take the simplest case that gauge
and gravitational anomalies only take place at the CH due to excluding the classically irrelevant outgoing modes, and disregard the quantum
contributions of the omitted ingoing modes at the EH.

\subsection{Hawking radiation from the CH}

Now we first determine the $U(1)$ gauge current flux and the $U(1)$ gauge anomaly at the CH. Similarly, as the $U(1)$ gauge anomaly takes place at the CH
due to integrating out the classically irrelevant outgoing modes, the $U(1)$ gauge current becomes anomalous near the CH $r_c-\epsilon \leq r
\leq r_c$, and obeys the anomalous equation  $\partial_r\mathcal{J}_{(C)}^r=-m^2 \partial_r \mathcal{A}_t/(4\pi)$ for the left-handed fields(ingoing modes). In the other
region, the current is conserved $\partial_r \mathcal{J}_{(o)}^r=0$. The total $U(1)$ gauge current can be written as a sum of the two regions, that
is $\mathcal{J}^r= \mathcal{J}_{(o)}^r\Theta_-+\mathcal{J}_{(C)}^r C$. Here, $\Theta_-=\Theta(r_c-r- \epsilon)$ and $C=1-\Theta_-$ are, respectively,
the scalar step function and top hat function. Under gauge transformation, the variance of the effective action (without the classically irrelevant
outgoing modes at the CH) can be written as
\bea
-\delta W &=& \int dtdr \lambda \Big[-\partial_r\Big(\frac{m^2}{4\pi}\mathcal{A}_t C\Big) \nn\\
&+&\Big(\mathcal{J}_{(C)}^r-\mathcal{J}_{(o)}^r+\frac{m^2}{4\pi}\mathcal{A}_t\Big)\delta(r-r_c+\epsilon)\Big]. \label{j1}
\eea
In (\ref{j1}), the first term is cancelled by the quantum effect of the classically irrelevant outgoing modes at the CH, whose contributions
to the total current are ${m^2}\mathcal{A}_t C/(4\pi)$. The effective action should be gauge invariance at the quantum level, which relates
\be
h_o=h_c+\frac{m^2}{4\pi}\mathcal{A}_t(r_c),
\ee
where $h_o$ is the $U(1)$ gauge current flux observed by an observer who lives between the EH and the CH, and $h_c$ is that at the CH.
Its values can be easily determined as
\bea
h_o &=& \mathcal{J}_{(o)}^r, \nn\\
h_c &=& \mathcal{J}_{(C)}^r+\frac{m^2}{4\pi}\int_{r_c}^r dr \partial_r \mathcal{A}_t.
\eea
Imposing the covariant current, which is related to the consistent one as $\widetilde{\mathcal{J}}^r=\mathcal{J}^r-m^2\mathcal{A}_t C/(4\pi)$, vanishes
at the CH, we can easily determine the value of the $U(1)$ gauge current flux
to be
\be
h_o=\frac{m^2}{2\pi}\mathcal{A}_t(r_c)=-\frac{\Xi m^2 a}{2\pi(r_c^2+a^2)}. \label{ho}
\ee
This gauge current flux corresponds to the angular momentum flux of Hawking radiation from the CH of the black hole. The negative sign denotes that
the effective field theory must absorb the gauge current flux to ensure gauge invariance at the quantum level.

Also, we can determine the total energy momentum tensor flux radiated from the CH of the Kerr-de Sitter black hole. In addition to gauge symmetry, the
effective two-dimensional theory for each partial wave has general coordinate symmetry. Here we take the simplest case discussed above.
Near the CH $r_c-\epsilon \leq r \leq r_c$, the effective field theory contains $U(1)$ gauge and gravitational anomalies, which gives a
great constraint on the energy momentum tensor
\be
\partial_rT_{t(C)}^r=\mathcal{J}_{(C)}^r\partial_r \mathcal{A}_t+ \mathcal{A}_t\partial_r \mathcal{J}_{(C)}^r-\partial_r\mathcal{N}_t^r.
\ee
In the other region, since there is a $U(1)$ gauge field background, the energy momentum tensor satisfies the modified conservation equation
$\partial_r T_ {t(o)}^r=h_o\partial_r  \mathcal{A}_t$. Under general coordinate transformation, the effective action (here we omit the
contributions of the classically irrelevant outgoing modes at the CH) changes as
\bea
-\delta W&=& \int dtdr \lambda^t \Big[h_o\partial_r \mathcal{A}_t-\partial_r\Big(\frac{m^2}{4\pi} \mathcal{A}_t^2+\mathcal{N}_t^r\Big)C \nn\\
&+&\Big(T_{t(C)}^r-T_{t(o)}^r+\frac{m^2}{4\pi} \mathcal{A}_t^2+\mathcal{N}_t^r\Big)\delta(r-r_c+\epsilon)\Big]. \label{51}
\eea
In (\ref{51}), the first term is the classical effect of the background gauge field for constant current flow. The second term is cancelled by the
quantum effect of the classically irrelevant outgoing modes at the CH, whose contributions to the total energy momentum tensor are $\Big({m^2}
\mathcal{A}_t^2/({4\pi})+\mathcal{N}_t^r\Big)C$. To restore general coordinate covariance at the quantum level, the coefficient of the delta
function should also vanish. So we have
\be
n_o=n_c-\frac{m^2}{4\pi}\mathcal{A}_t(r_c)+\mathcal{N}_t^r(r_c),
\ee
where
\bea
n_o&=&T_{t(o)}^r-h_o\mathcal{A}_t(r), \nn\\
n_c &=&T_{t(C)}^r-\int_{r_c}^{r}dr\partial_r\Big[h_o\mathcal{A}_t-\frac{m^2}{4\pi}\mathcal{A}_t^2-\mathcal{N}_t^r\Big],
\eea
are the energy flow observed by an observer who lives between the EH and the CH, and that at the CH, respectively. To assure the regularity of
the physical quantities, we impose the covariant energy momentum tensor, which is related to the consistent one as described by Eq.(\ref{11}),
vanishes at the CH, that condition determines the total energy momentum tensor flux as
\be
n_o=-\frac{\Xi^2 m^2 a^2}{4\pi(r_c^2+a^2)^2}-\frac{\pi}{12}T_c^2. \label{no}
\ee
In (\ref{no}), $T_c=-\partial_rf(r)/(4\pi)|_{r=r_c}$ is the Hawking temperature at the CH of the black hole. The negative sign denotes that the
energy momentum tensor flux is radiated into the effective theory to ensure general coordinate covariance at the quantum level.

In fact, these absorbing gauge current and energy momentum tensor fluxes required to restore gauge invariance and general coordinate covariance at the
the quantum level, and respectively expressed by Eq.(\ref{ho}) and Eq.(\ref{no}), are exactly equal to those of blackbody radiation moving in the
negative $r$ direction with the Hawking distribution at the CH. For fermions, the Hawking distribution at the CH formally takes the form as
$ \mathcal{N}_{\pm m}(\omega)=-1/[\exp(\frac{\omega\pm mA_t(r_c)}{T_c})+1]$ (here the negative sign denotes Hawking radiation is radiated
into the black hole from the CH). Integrating this distribution, the angular momentum and energy momentum tensor fluxes at the CH can be
obtained to take the same form as Eq.(\ref{ho}) and Eq.(\ref{no}), respectively.

\section{Conclusions and Discussions}\label{cm}

In this paper, we study Hawking radiation of different-type black holes in de Sitter spaces via anomalous point of view, specifically
including that of the purely de Sitter black hole, and the static, spherically symmetric Schwarzschild-de Sitter black hole as well as the
rotating Kerr-de Sitter black hole. These black holes shares in common to have a de Sitter cosmological horizon(CH). At the CH, the outgoing
modes that fall out of the de Sitter cosmological horizon(CH) would never fall classically back since the cosmological horizon(CH) is a null
hypersurface, but Quantum mechanically its contributions on the physics inside the cosmological horizon(CH) should be taken into account.
In the two-dimensional reduction, when the effective field theory is formulated to exclude the classically irrelevant outgoing modes at
the CH, it is chiral here, but contains gauge or gravitational anomaly. In order to cancel these anomalies and restore gauge invariance
or general coordinate covariance at the quantum level, each partial wave of the scalar field must be in a state with a net gauge current
flux given by Eq.(\ref{ho}) and energy momentum tensor flux given by Eq.(\ref{ao2}), Eq.(\ref{fo}) and Eq.(\ref{no}). The result shows that
these absorbing fluxes are exactly equal to those of Hawking radiation from the de Sitter cosmological horizon(CH).

In the case of black holes with a repulsive $\Lambda$ term, there are an event horizon(EH) and a de Sitter cosmological horizon(CH) for
an observer who moves on the world line of constant $r$ between the EH and the CH. The effective field theory that describes an
observable physics should then be formulated between the EH and the CH to, respectively, integrate out the classically irrelevant ingoing modes
at the EH and the classically irrelevant outgoing modes at the CH. To simplify our discussion, we have treated the EH and the CH as two
independent systems. That is to say, when dealing with Hawking radiation from the EH, we have overlooked the effect coming from
the CH. Similarly, the effect coming from the EH is ignored when we consider the de Sitter radiation from the CH. This simplification
can be seen in Ref.\cite{MJ} to derive Hawking radiation via tunnelling from black holes in de Sitter spaces. The result shows that
Hawking fluxes from the black hole horizon(EH) and the de Sitter cosmological horizon(CH) are capable of cancelling gauge or
gravitational anomaly and restoring gauge invariance or general coordinate covariance at the quantum level.

In fact, the Robinson-Wilczek's method can be universally extended to derive Hawking fluxes from any static or
stationary black hole, which takes the form as $ds^2=f(r)dt^2-g^{-1}(r)dr^2$ after a dimensional reduction technique
near the horizon. In our discussion, $f(r)=g(r)$ and the determinant of its diagonal metric is the unity. If its value
differs from the unity, namely $f(r)\neq g(r)$, the Robinson-Wilczek's derivation of Hawking radiation via anomalies can also
be applicable, but some formula need to be modified by $\sqrt{f(r)/g(r)}$(specifically see in Ref.\cite{ALL6}).

{\bf Acknowledgments}: ~This work was partially supported by the Natural Science Foundation of
China under Grant No. 10675051, 10635020, 70571027, and a grant by the Ministry of Education of
China under Grant No. 306022.


\begin{thebibliography}{99}

\bibitem{SWH}
S. Hawking, {\it Particle creation by black holes}, {\it Commun. Math. Phys.} \textbf{43} (1975) 199.


\bibitem{GH}
G. Gibbons and S. Hawking, {\it Action integrals and partition functions in quantum gravity},   {\it Phys. Rev. D} \textbf{15} (1977) 2752.

\bibitem{SC}
A. Strominger and C. Vafa, {\it Microscopic origin of the Bekenstein-Hawking entropy},  {\it Phys. Lett. B} \textbf{379} (1996) 99 [hep-th/9601029];

A. Peet, {\it TASI lectures on black holes in string theory}, [hep-th/0008241].

\bibitem{PW1}
M. Parikh and F. Wilczek, {\it Hawking Radiation As Tunneling},  {\it Phys. Rev. Lett.} \textbf{85} (2000) 5042 [hep-th/9907001];

P. Kraus and F. Wilczek, {\it Self-interaction correction to black hole radiance},  {\it Nucl. Phys. B} \textbf{433} (1995) 403 [gr-qc/9408003];
{\it Effect of self-interaction on charged black hole radiance},  {\it Nucl. Phys. B} \textbf{437} (1995) 231 [hep-th/9411219];

S. Hemming and E. Keski-Vakkuri, {\it Hawking radiation from AdS black holes},  {\it Phys. Rev. D} \textbf{64} (2001) 044006 [gr-qc/0005115];

E. C. Vagenas, {\it Are extremal 2D black holes really frozen?},  {\it Phys. Lett. B}  \textbf{503} (2001) 399 [hep-th/0012134]; {\it Generalization of
the KKW analysis for black hole radiation},  {\it Phys. Lett. B} \textbf{559} (2003) 65 [hep-th/0209185];

M. Arzano, A.J.M. Medved and E.C. Vagenas, {\it Hawking radiation as tunneling through the
quantum horizon},  {\it J. High. Energy. Phys} \textbf{09} (2005) 037 [hep-th/0505266];

J.Y. Zhang and Z. Zhao, {\it Hawking radiation of charged particles via tunneling from the
Reissner-Nordstrom black holes},  {\it J. High. Energy. Phys} \textbf{10} (2005) 055; {\it Massive particles¡¯ black hole tunneling
and de Sitter tunneling},  {\it Nucl. Phys. B} \textbf{725} (2005) 173;

W.B. Liu, {\it New coordinates for BTZ black hole and Hawking radiation via tunnelling},  {\it Phys. Lett. B} \textbf{634} (2006) 541 [gr-qc/0512099].

\bibitem{PW2}
E. C. Vagenas, {\it Complex paths and covariance of Hawking radiation in 2-D stringy black holes}, {\it Nuovo Cim.B} \textbf{117} (2002) 899 [hep-th/0111047];

K. Srinivasan and T. Padmanabhan, {\it Particle production and complex path analysis},   {\it Phys. Rev. D} \textbf{60} (1999) 024007 [gr-qc/9812028];

S. Shankaranarayanan, T. Padmanabhan, and K. Srinivasan, {\it Hawking radiation in different coordinate settings: complex paths approach},
 {\it Class. Quantum. Grav.} \textbf{19} (2002) 2671 [qr-qc/0010042];

M. Angheben, M. Nadalini, L. Vanzo, and S. Zerbini, {\it Hawking radiation as tunneling for extremal and rotating black holes},
 {\it J. High. Energy. Phys.} \textbf{05} (2005) 014 [hep-th/0503081].

\bibitem{MJ}
A.J.M. Medved, {\it Radiation via tunneling from a de Sitter cosmological horizon},
 {\it Phys. Rev. D} \textbf{66} (2002) 124009 [hep-th/0207247];

Q.Q. Jiang and S.Q. Wu, {\it Hawking radiation of charged particles as tunneling from Reissner-Nordstrom-de Sitter black holes with a global monopole},
{\it Phys. Lett. B} \textbf{635} 151 [hep-th/0511123].


\bibitem{RW}
S.P. Robinson and F. Wilczek, {\it Relationship between Hawking Radiation and Gravitational Anomalies},  {\it Phys. Rev. Lett.}
\textbf{95} (2005) 011303 [gr-qc/0502074].


\bibitem{IUW1}
S. Iso, H. Umetsu, and F. Wilczek, {\it Hawking Radiation from Charged Black Holes via Gauge and Gravitational Anomalies},
 {\it Phys. Rev. Lett.} \textbf{96} (2006) 151302 [hep-th/0602146].


\bibitem{IUW2}
S. Iso, H. Umetsu, and F. Wilczek, {\it Anomalies, Hawking Radiations and Regularity in Rotating Black Holes},  {\it Phys. Rev. D} \textbf{74}
(2006) 044017 [hep-th/0606018].


\bibitem{ALL1}
K. Murata and J. Soda, {\it Hawking radiation from rotating black holes and gravitational anomalies},  {\it Phys. Rev. D} \textbf{74}
(2006) 044018 [hep-th/0606069].

\bibitem{ALL2}
E.C. Vagenas and S. Das, {\it Gravitational anomalies, Hawking radiation, and spherically symmetric black holes},
 {\it J. High. Energy. Phys.} \textbf{10} (2006) 025 [hep-th/0606077].

\bibitem{ALL3}
Z. Xu and B. Chen, {\it Hawking radiation from general Kerr-(anti)de Sitter black holes},  {\it Phys. Rev. D} \textbf{75} (2007) 024041 [hep-th/0612261].


\bibitem{ALL4}
M.R. Setare, {\it Gauge and gravitational anomalies and Hawking radiation of rotating BTZ black holes},  {\it Euro. Phys. J. C}  \textbf{49}
(2006) 865 [hep-th/0608080].

\bibitem{ALL5}
S. Iso, T. Morita, and H. Umetsu, {\it Quantum anomalies at horizon and Hawking
radiations in Myers-Perry black holes}, {\it J. High. Energy. Phys.} \textbf{04} (2007) 068 [hep-th/0612286]; {\it Higher-spin
Currents and Thermal Flux from Hawking Radiation},  {\it Phys. Rev. D} \textbf{75} (2007) 124004 [hep-th/0701272].

Q.Q. Jiang, S.Q. Wu, and X. Cai, {\it Hawking radiation from the (2+1)-dimensional BTZ black holes}, {\it Phys. Lett. B} \textbf{651} (2007) 58 [hep-th/0701048];

K. Xiao, W.B. Liu and H.B. Zhang, {\it Anomalies of the Achucarro-Ortiz black hole},  {\it Phys. Lett. B} \textbf{647} (2007) 482 [hep-th/0702199].

\bibitem{ALL6}
Q.Q. Jiang, S.Q. Wu, and X. Cai, {\it Hawking radiation from dilatonic black holes via anomalies}, {\it Phys. Rev. D} \textbf{75}
(2007) 064029; Erratum-ibid.\textbf{76} (2007) 029904 [hep-th/0701235].

S.Q. Wu and J.J. Peng, {\it Anomalies and Hawking radiation from the Reissner-Nordstrom black hole with a global monopole}, [0706.0983][hep-th].

\bibitem{ALL7}
Q.Q. Jiang and S.Q. Wu, {\it Hawking radiation from rotating black holes in anti-de Sitter spaces via gauge and gravitational anomalies},
{\it Phys. Lett. B} \textbf{647} (2007) 200 [hep-th/0701002].

\bibitem{HJ}
H. Shin and W. Kim, {\it Hawking Radiation from Non-Extremal D1-D5 Black Hole via Anomalies}, {\it J. High. Energy. Phys.} \textbf{06} (2007) 012 [0705.0265][hep-th];

J.J. Peng and S.Q. Wu,  {\it Hawking radiation from the Schwarzschild black hole with a global monopole via gravitational anomaly}, [0705.1225][hep-th];

S. Das, S.P. Robinson and E.C. Vagenas, {\it Gravitational anomalies: a recipe for Hawking radiation}, [0705.2233][hep-th].


\bibitem{ASJ}
A.G. Riess et al, {\it Observational evidence from supernovae for an accelerating universe and a cosmological constant,} {\it Astron. J}
\textbf{116} (1998) 1009 [astro-ph/9805201];

S. Perlmutter et al, {\it Measurements of $\Omega$ and $\Lambda$ from 42 High-Redshift Supernovae}, {\it Astrophys. J} \textbf{517}
(1999) 565 [astro-ph/9812133];

J.P. Ostriker, P.J. Steinhardt, {\it The observational case for a low-density Universe with a non-zero cosmological constant},
{\it Nature} \textbf{377} (1995) 600.

\bibitem{AD}
M. Park, {\it Statistical entropy of three-dimensional Kerr-de Sitter space,} {\it Phys. Lett. B}, \textbf{440} (1998) 275 [hep-th/9806119];
{\it Symmetry algebras in Chern-Simons theories with boundary: canonical approach}, {\it Nucl. Phys. B} \textbf{544} (1999) 377 [hep-th/9811033].


A. Strominger, {\it The dS/CFT correspondence,} {\it J. High. Energy. Phys.} \textbf{10} (2001) 034 [hep-th/0106113];

D. Klemm, {\it Some aspects of the de Sitter/CFT correspondence,} {\it Nucl. Phys. B} \textbf{625} (2002) 295 [hep-th/0106247].


\bibitem{LE}
L. Alvarez-Gaume and E. Witten, {\it Gravitational anomalies,} {\it Nucl. Phys. B} \textbf{234}, 269 (1984);

R. Bertlmann and E. Kohlprath, {\it Two-Dimensional Gravitational Anomalies, Schwinger Terms, and Dispersion Relations,}
{\it Ann. Phys. (N.Y.)} \textbf{288} (2001) 137 [hep-th/0011067].

\bibitem{GU}
G. W. Gibbons and S. W. Hawking,  {\it Cosmological event horizons, thermodynamics, and particle creation,}  {\it Phys. Rev. D} \textbf{16} (1977) 2738;

U. Khanal, {\it Rotating black hole in asymptotic de Sitter space: Perturbation of the space-time with spin fields,}
{\it Phys. Rev. D} \textbf{28} (1983) 1291.

\end{thebibliography}
\end{document}